\documentclass[%
 apl,
 amsmath,amssymb,
 reprint,%
]{revtex4-1}

\usepackage{graphicx}
\usepackage{dcolumn}
\usepackage{bm}

\usepackage[utf8]{inputenc}
\usepackage[T1]{fontenc}
\usepackage{mathptmx}
\usepackage{array}
\usepackage{booktabs}
\usepackage{xcolor,colortbl}

\definecolor{VTTgray}{rgb}{0.82,0.85,0.87}
\definecolor{VTTgreen}{rgb}{0.42,0.70,0.53}
\definecolor{VTTblue}{rgb}{0,0.53,0.68}

\newcolumntype{a}{>{\columncolor{VTTgray}}c}
\newcolumntype{b}{>{\columncolor{white}}c}

\usepackage{textcomp}
\newcommand{\murm}{\hbox{\textmu}}

\begin{document}


\title[maintitle]{Cascaded superconducting junction refrigerators: optimization and performance limits}

\author{A. Kemppinen}
\email{antti.kemppinen@vtt.fi}
\author{A. Ronzani}%
\author{E. Mykk\"anen}%
\author{J. H\"atinen}%
\author{J. S. Lehtinen}%
\author{M. Prunnila}%
\affiliation{VTT Technical Research Centre of Finland Ltd, 02150 Espoo, Finland}

\date{\today}

\begin{abstract}
We demonstrate highly transparent silicon--vanadium and silicon--aluminum tunnel junctions with relatively low sub-gap leakage current and discuss how a tradeoff typically encountered between transparency and leakage affects their refrigeration performance. We theoretically investigate cascaded superconducting tunnel junction refrigerators with two or more refrigeration stages. In particular, we develop an approximate method that takes into account self-heating effects, but still allows to optimize the cascade a single stage at a time. We design a cascade consisting of energy-efficient refrigeration stages, which makes cooling of, e.g., quantum devices from above 1~K to below 100~mK a realistic experimental target.
\end{abstract}

\maketitle

Normal metal -- insulator -- superconductor (NIS) and semiconductor -- superconductor (Sm--S) tunnel junctions can be used for cryogenic electrical refrigeration, because the superconducting energy gap $\Delta$ and a bias voltage $V\lesssim \Delta/e$ allow thermionic energy filtering of the tunneling electrons~\cite{Nahum1994, Leivo1996,Savin2001,Giazotto2006}. Here $e$ is the elementary charge. Suspended lateral cold finger assemblies have been used to refrigerate electrically both
 electrons and phonons of macroscopic objects in proof-of-concept experiments~\cite{Lowell2013,Miller2008,Clark2005}. Despite extensive efforts and potential for low-cost, compact, and maintenance-free devices, electrical refrigerators have not replaced conventional techniques such as dilution refrigeration. Challenges for practical applications include limited cooling power and complicated engineering of phonon and electron--phonon heat flows~\cite{Rajauria2007,Koppinen2009,Muhonen2009,Nguyen2015}, and the limited temperature range for refrigeration that depends on $\Delta$. The latter could in principle be overcome by multi-stage coolers that utilize superconductors with different $\Delta$~\cite{Quaranta2011,Nguyen2016,Camarasa2014}.

\begin{figure}
\includegraphics[width=0.5\textwidth]{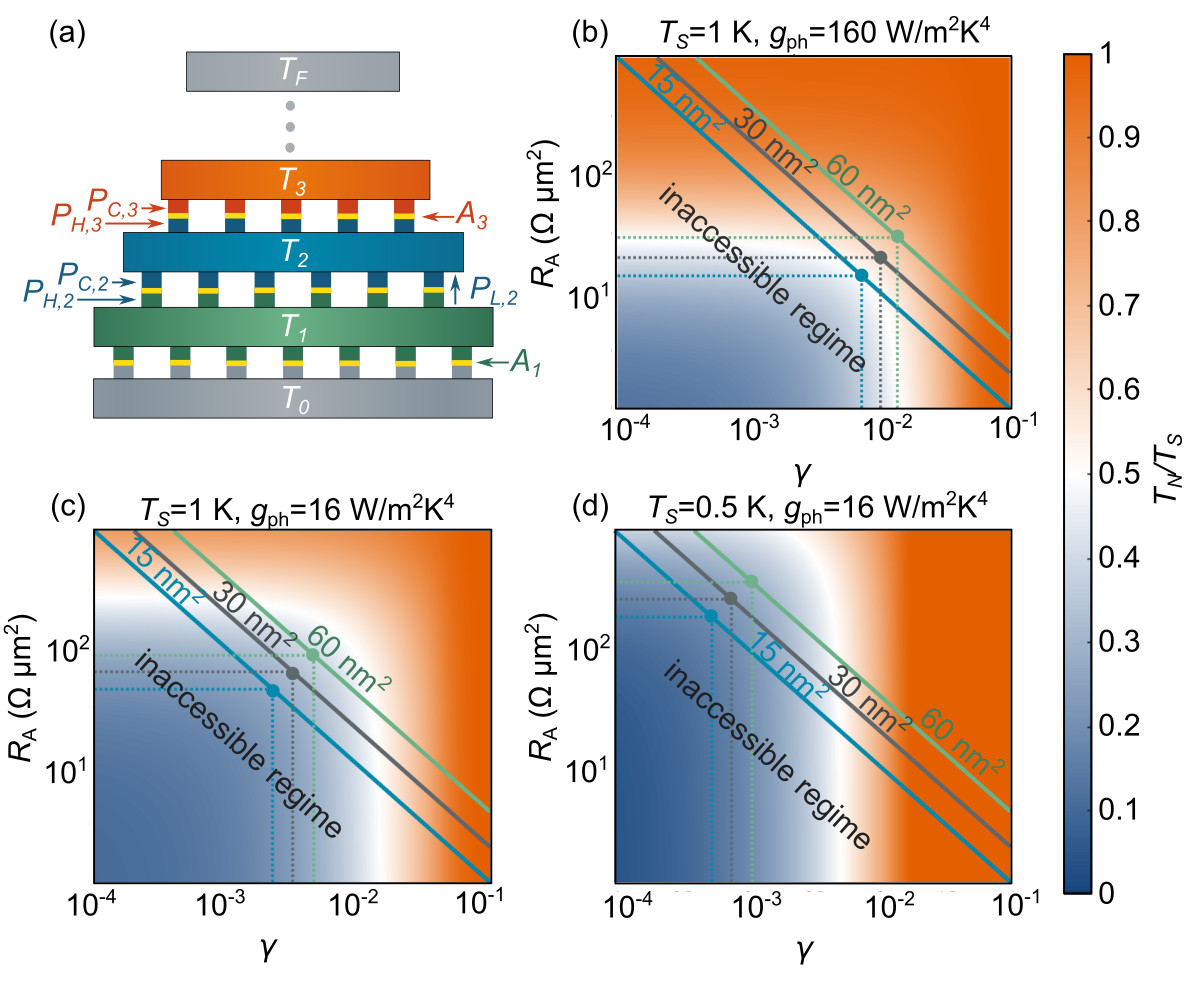}
\caption{\label{fig:gammara} (a) Schematic picture of a cascaded refrigerator with $F$ (individually biased) stages. A conventional refrigerator provides the base temperature $T_0$. 
Each stage $n$ has tunnel junctions with total area $A_n$ and a macroscopic volume to thermalize electrons and phonons to a single temperature $T_n$. The thermal balance of stage $n$ is defined by the cooling power $P_{C,n}$, phonon heat leak $P_{L,n}$ from stage $n-1$, and the heating power $P_{H,n+1}$ resulting from the refrigeration of stage $n+1$.
 (b-d) Simulated $T_N/T_S$ as a function of $\gamma$ and $R_A$ at for V--Si junctions. The three solid lines indicate the limiting trade-off between $R_A$ and $\gamma$ for $A_\gamma=15, 30$, and 60~nm$^2$ from left to right, respectively. The circles and dotted lines indicate $\gamma$ and $R_A$ that yield lowest $T_N/T_S$ on each line. Panels (b-c) are simulated at $T_S=1$~K, and panel (d) at $T_S=0.5$~K. Panel (b) is simulated with $g_\mathrm{ph}=160~\mathrm{W/m^2K^4}$ and panels (c--d) with $g_\mathrm{ph}=16~\mathrm{W/m^2K^4}$.}
\end{figure}

Recently, the electronic refrigeration of a macroscopic silicon chip was demonstrated using Sm--S (Al--Si) junctions~\cite{Mykkanen2020}. They were used as the cooling element and mechanical support, and as
a blockade for phonon heat transport based on the Kapitza resistance~\cite{Swartz1987} between Al and Si.
This approach avoids complex arrangements of cold fingers, which should allow a simple multi-stage assembly, possibly even 3D integration, see Fig.~\ref{fig:gammara}(a).
Reference \cite{Mykkanen2020} concludes that refrigeration from above 1~K to below 100~mK is a realistic target, but requires significant further development: \emph{(i)} 
Since both cooling power and phonon heat leaks are proportional to the tunnel junction area $A$, it is beneficial to decrease the characteristic resistance
$R_A=R_TA$ to improve the ratio between cooling power and phonon heat leaks. Here, $R_T$ is the tunneling resistance of the junction.
\emph{(ii)} Suppression of phonon thermal conductance at temperatures above 500~mK requires other phonon blocking methods in addition to the Kapitza resistance, e.g., nanowire constrictions~\cite{Mykkanen2020a}. \emph{(iii)} A superconductor with larger $\Delta$ than aluminum, e.g., vanadium, is needed for refrigeration above 
about 500~mK.

In this letter, we demonstrate both Si--Al and Si--V tunnel junctions with small $R_A$ and discuss the optimization of a cascaded cooler with multiple refrigeration stages. In particular, we evaluate how the thermal balance of the device depends on finite cooling efficiency, i.e., that cooling one side of a junction with power $P_C$ causes heating power $P_H$ on the other side. We use Fig.~\ref{fig:gammara}(a) as our model system, but our method is applicable also for more complex designs.

An important limitation for cooling with NIS or Sm--S tunnel junctions, especially at temperatures well below the critical temperature of the superconductor $T\ll T_c$, is the leakage resistance $R_0$ in the sub-gap regime. The sub-gap leakage can originate from nonidealities of the superconductor or the tunnel junction, but the fundamental limit is set by Andreev reflection~\cite{Gunnarsson2015,Andreev,Rajauria2008}. According to theory for opaque NIS junctions~\cite{Maisi}, the sub-gap leakage parameter $\gamma=R_0/R_T$ due to Andreev reflection is inversely proportional to the characteristic resistance: $\gamma=A_\mathrm{ch}R_K/(4R_A)$. Here, $A_\mathrm{ch}$ is the effective tunneling area per conductance quantum, $R_K=h/e^2$ is the quantum resistance, and $h$ is the Planck constant. Andreev reflection thus prevents having low $\gamma$ and $R_A$ simultaneously. The observed $A_\mathrm{ch}$ of Andreev limited Cu--AlO--Al junctions is between about 10\ldots 30~nm$^2$, factor 5\ldots 10 higher than theoretical estimates, which has been explained by inhomogeneities of the tunnel barrier~\cite{Eiles1993,Pothier1994,Maisi,Marin2020}. There is no reference value or theoretical prediction for $A_\mathrm{ch}$ in Sm--S junctions, which possibly depends on the microscopic quality of tunnel junctions. Therefore we define a phenomenological leakage area,
\begin{equation} \label{eq:andreev}
 A_\gamma=  \frac{ 4\gamma  R_A}{R_K },
\end{equation}
and use it as a figure of merit for tunnel junctions. Parameter $A_\gamma$ yields an upper limit for $A_\mathrm{ch}$ and allows empiric means for comparing leakage strength in junctions with different $R_A$ in view of the fundamental trade-off between $\gamma$ and $R_A$.

Figure~\ref{fig:gammara}(b) shows the simulated relative cooling $T_N/T_S$, i.e., the ratio between temperatures of the Sm and S sides for a Si--V junction at $T_S=1$~K and with superconducting gap is $\Delta_\mathrm{V}=600$~\murm{}eV~\cite{Garcia2009} ($T_c \approx 4$~K). The thermal power of phonon leakage that limits cooling is
\begin{equation}\label{eq:pl}
P_L=g_\mathrm{ph}A(T_S^4 - T_N^4),
\end{equation}
which is valid for several processes, but the prefactor $g_\mathrm{ph}=160$~W$/$m$^2$K$^4$ has been measured for the Kapitza resistance of Al--Si interface~\cite{Mykkanen2020,Swartz1987}. Figures~\ref{fig:gammara}(c--d) show similar simulations at 1~K and 0.5~K, respectively, where $g_\mathrm{ph}=16$~W$/$m$^2$K$^4$, obtained, e.g., with phonon engineering techniques~\cite{Mykkanen2020}. We illustrate the limiting tradeoff between $\gamma$ and $R_A$ for 3 example values of $A_\gamma$. Lower $R_A$ at cost of higher $\gamma$ is preferred when $P_L$ is high, i.e., for large $g_\mathrm{ph}$ and $T_S$. Smaller $g_\mathrm{ph}$ or $T_S$ shift the preference to higher $R_A$ with lower $\gamma$, but overall Figs.~\ref{fig:gammara}(b--d) motivate the fabrication of tunnel junctions with $R_A\leq100$~$\Omega$\murm{}m$^2$.

\begin{figure}
\includegraphics[width=0.5\textwidth]{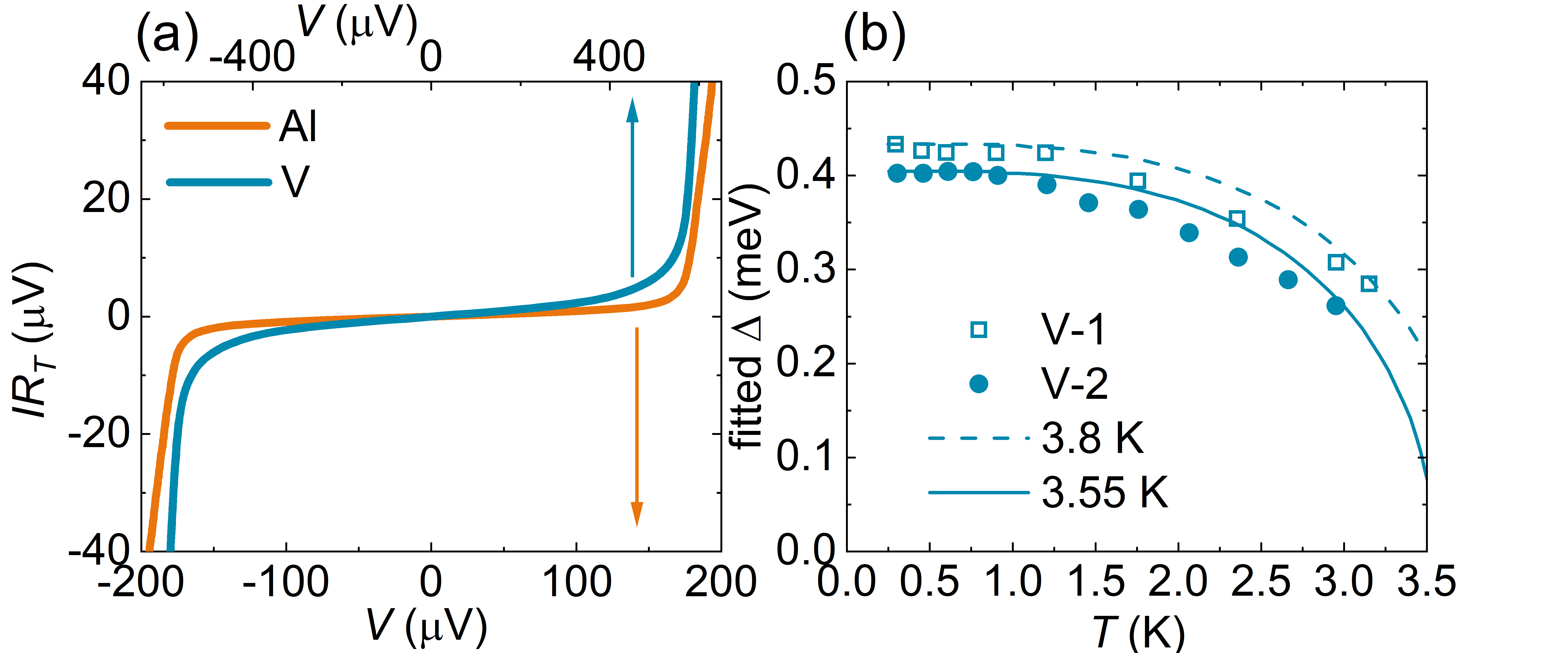}
\caption{\label{fig:junctions} (a) Current--voltage characteristics of transparent Si--Al (orange) and Si--V (blue) junctions. Both junctions were measured in series with a similar junction with significantly larger area, which does not affect the measurement of $\gamma$.
 (b) Estimated $\Delta$ of two Si--V junctions as a function temperature. These effective values are obtained from fitting a model of two junctions with different areas to the current–voltage characteristics (circles and squares). The lines are a comparison with Bardeen-Cooper-Schrieffer (BCS) theory with $\Delta$ corresponding to $T_c$ of 3.55~K and 3.8~K,  rescaled by empirical factor of 0.75 to match the gap estimates at $T=0$~K. A discrepancy with BCS theory is expected because the vanadium side is actually a stack of V and Al.}
\end{figure}

Figure~\ref{fig:junctions}(a) shows the voltage--current characteristics of Si--Al and an Si--V junctions fabricated with a similar process as in Ref.~\cite{Gunnarsson2015}, but with increased junction transparency. The Si--Al junctions have 500 nm thick Al as superconductor, whereas Si--V junctions have a multilayer stack Si -- SiO$_x$ -- Al (25 nm) -- V (150 nm) -- Al (400 nm). High-quality vanadium junctions require a thin layer of, e.g., Al, at the junction interface~\cite{Quaranta2011}. Changing the thicknesses of V and the interface layer allows tuning $\Delta$ roughly between those of Al and V, $\Delta_\mathrm{Al}=200$~\murm{}eV and $\Delta_\mathrm{V}=600$~\murm{}eV, respectively. The thicker Al layer reliefs the multilayer film strain and improves quasiparticle diffusion due to its lower $\Delta$ and normal-state resistivity compared to V. The Si--Al junction has $R_A=48$~$\Omega\,$\murm{}m$^2$, which lower by factor 10 than the results of Ref.~\cite{Mykkanen2020}. Yet it still has a relatively low $\gamma=8\times10^{-3}$, yielding $A_\gamma=62$~nm$^2$. The same parameters for the Si--V junction are $R_A= 71$~$\Omega\,$\murm{}m$^2$, $\gamma=6\times10^{-3}$, and $A_\gamma=65$~nm$^2$. These are significant improvements to previous results with Sm--S junctions~\cite{Mykkanen2020,Gunnarsson2015}. Notably, typical experiments of NIS junctions have about an order of magnitude higher $R_A$, which may explain correspondingly lower values of $\gamma$~\cite{Maisi,Rajauria2008}.

Literature on Sm--S or NIS coolers usually aims at minimizing $T_N/T_S$ as in Fig.~\ref{fig:gammara}, but also energy efficiency is crucial for a practical refrigerator.
In the cascade of Fig.~\ref{fig:gammara}(a), the heating power $P_{H,n}$, caused by cooling stage $n$, must not overload the subsequent stage $n-1$. Unfortunately, the numerical optimization of the cascade is computationally demanding due to the large number of parameters that affect its performance. Neither does that provide physical insight into the system. Therefore, we develop an approximate method for optimizing the cascade a single stage at a time.

The heat currents to the normal (N) and superconducting (S) sides of an Sm--S or NIS junction and for tunneling to ($+$) and from ($-$) the normal side are \cite{Giazotto2006}
\begin{align} \label{eq:qint}
\dot{Q}_N^\pm & =\pm\frac{A\Delta^2}{eR_A}\int_{-\infty}^{\infty}\mathrm{d}\epsilon n_S(\epsilon ,\gamma)f_S(\pm\epsilon,\tau_S)f_N(\pm v\mp\epsilon,\tau_N)  (\epsilon -v), \nonumber \\
\dot{Q}_S^\pm & =\pm\frac{A\Delta^2}{eR_A}\int_{-\infty}^{\infty}\mathrm{d}\epsilon n_S(\epsilon ,\gamma)f_S(\pm\epsilon,\tau_S)f_N(\pm v\mp\epsilon,\tau_N) \epsilon.
\end{align}
Here, $\epsilon=E/\Delta$ and $v=eV/\Delta$ are normalized energy and voltage, respectively, $f_N$ and $f_S$ are Fermi functions at normalized temperatures $\tau_i=T_i/\Delta$. We use the Dynes density of states ($n_S$) to describe $R_0$~\cite{Dynes1984}.
Ideally, for any target temperature $T_N$ we keep the integral parts of Eqs.~(\ref{eq:qint}) constant, in their optimal value, by using fixed $v$ and by choosing an optimal superconductor with
$\Delta_\mathrm{opt}\propto T_N,T_S$~\cite{Mykkanen2020} that fixes $\tau_i$. This is possible, e.g., by utilizing the proximity effect. Note that Eqs.~(\ref{eq:qint}) are insensitive to $T_S$ when $T_S\ll T_c$ as then the quasiparticle density is small. Therefore, when $\Delta$ is scaled with $T_N$, both $P_C=-\dot{Q}_N^+-\dot{Q}_N^-$ and $P_H=\dot{Q}_S^++\dot{Q}_S^-$ scale as $\propto T_N^2$.  This suggests that if a refrigeration stage yields a temperature ratio $T_N^2/T_S^2\lesssim P_C/P_H$, the subsequent stage of a cascade that provides temperature $T_S$ may have sufficient cooling power to mitigate $P_H$, see Fig.~\ref{fig:gammara}(a). If the refrigeration performance does not satisfy this condition, excessive heating can be compensated by scaling tunnel junction areas, i.e., by having $A_1>A_2>A_3\ldots$  because the cooling and heating power of stage $n$ are proportional to its area, i.e., $P_{C,n}, P_{H,n} \propto A_n$.

The maximum heat load tolerated by a refrigeration stage, $P_{\mathrm{in}}=P_{C}-P_{L}$ also depends on $P_{L}$. Unfortunately, $P_L$ of Eq.~(\ref{eq:pl}) follows a $T^4$ power law and increases faster as function of temperature than $P_C$. Thus cooling becomes increasingly difficult at higher temperatures even if materials with suitable $\Delta$ are available. However, thermal balance adjusts $T_N$ so that $P_C\geq P_L$. Hence we expect that the proportionality $P_C\propto T_N^2$ will approximately determine the magnitude of cooling power available for external load so that $P_{\mathrm{in}}\propto T_N^2$. Therefore, when stage $n$ is the dominant heat load to stage $n-1$, i.e., $P_{H,n}=P_{\mathrm{in},n-1}$, heating of stage $n$ can be compensated with the ratio of areas
\begin{equation}
\label{eq:areas}
a_{n} = \frac{A_{n-1}}{A_n} \approx\frac{P_{\mathrm{in},n-1}T_n^2}{P_{\mathrm{in},n}T_{n-1}^2}= \frac{P_{H,n}T_n^2}{P_{\mathrm{in},n}T_{n-1}^2}
\equiv \frac{P_HT_N^2}{P_\mathrm{in}T_S^2}=a.
\end{equation}
The right hand side is written as the function of parameters of a single stage, i.e., this requirement allows to consider the performance of a single stage at a time.

An optimal refrigerator reaches the target temperature $T_F$ from $T_0$ with minimum $P_{H,1}/P_{\mathrm{in},F}$, i.e., it maximizes the cooling power with respect to
the heating power. Since cooling and heating powers are proportional to the tunnel junction areas, our goal is to minimize $\prod_{n=1}^Fa_{n}$. We do not know
\emph{a priori} whether it is optimal to have many stages (large $F$) with minimal $a$ or, conversely, small $F$ with larger $a$. Therefore we find an approximate optimum by considering $F$ stages that are equal on the logarithmic scale, i.e., have the same $a$ and $T_N/T_S$. Then  $T_0/T_F=(T_S/T_N)^F$ and $F=\lg(T_0/T_F)/\lg(T_S/T_N)$, which yields
$\prod_{n=1}^Fa_{n}=a^F=O^{\lg(T_0/T_F)}$, where $O$ is an optimization parameter we aim to minimize:
\begin{equation}
\label{eq:odef}
O \equiv a^\frac{1}{\lg (T_S/T_N)} = \left( \frac{P_HT_N^2}{P_\mathrm{in}T_S^2} \right) ^\frac{1}{\lg (T_S/T_N)}.
\end{equation}
Note that $O$ summarizes the performance of each individual stage based on physical quantities (power and temperature) that characterize it. Therefore $O$ allows to optimize refrigeration stages independently of others. There is no direct dependence on phenomenological modeling parameters such as $\gamma$ or $g_\mathrm{ph}$ and a low $O$ may be obtained in several ways. Parameter $O$ is a more comprehensive figure of merit compared to the typical optimization of $T_N/T_S$ shown in Figs.~\ref{fig:gammara}(b--d) since it also takes into account the energy efficiency. To gain intuitive insight to the magnitude of $O$, let us consider a refrigerator with $T_1=1$~K and $T_F=100$~mK. If stages $n=2,\ldots ,F$ have the same $O=O_n$, the cascade then requires $A_1/A_F=O^{\lg (1\; \mathrm{K}\, /\, 100\; \mathrm{mK})}=O$.

If the contribution of $P_L$ is small, for example because of good phonon filtering or low $T_S$, then $P_{H,n}$ dominates the heat balance of stage $n-1$, i.e., $P_{H,n}\approx P_{C,n-1}$, which yields the scaling $a^\dagger= P_H T_N^2/(P_C T_S^2)$ instead of Eq.~(\ref{eq:areas}) and $O^\dagger\equiv (a^\dagger)^{1/\lg (T_S/T_N)}$ instead of Eq.~(\ref{eq:odef}). 

Next, we study by simulations, how $v$, $\gamma$, $R_A$, $g_\mathrm{ph}$, $\Delta$, $T_S$, and $T_N$ affect $O$ for Si--Al and Si--V junctions. We use Eq.~(\ref{eq:andreev}) and a targeted value $A_\gamma=15$~nm$^2$ to calculate $R_A(\gamma)$, which eliminates one dimension of the optimization problem. We focus on 3 specific values of $\Delta$, motivated by our experiments: 200, 450, and 600~\murm{}eV. This approach requires computing $P_C$, $P_H$, and $P_L$ in 5-dimensional space ($v$, $\gamma$, $g_\mathrm{ph}$, $T_S$, $T_N$) for each $\Delta$ separately, which is computationally demanding, but dramatically easier than varying all parameters of the full cascade. More dimensions can be taken into account when optimising $O$, but here we focus on what we consider as the simplest realistic model. For each set of $v$, $\gamma$, $g_\mathrm{ph}$, and $T_S$ we first find the balance temperature $T_{N,b}$ where $P_C=P_L$, and then temperature $T_{N,O}$ in range $T_{N,b}\leq T_{N,O} \leq T_S$ that yields minimum $O$ (alternatively $O^\dagger$). This computation allows us to determine, e.g., the requirements of efficient refrigeration for each value of $\Delta$.  Finally, we discuss the optimization of a specific 3-stage refrigerator and compare our approximate results to simulations of the complete thermal balance.

\begin{figure}
\includegraphics[width=0.5\textwidth]{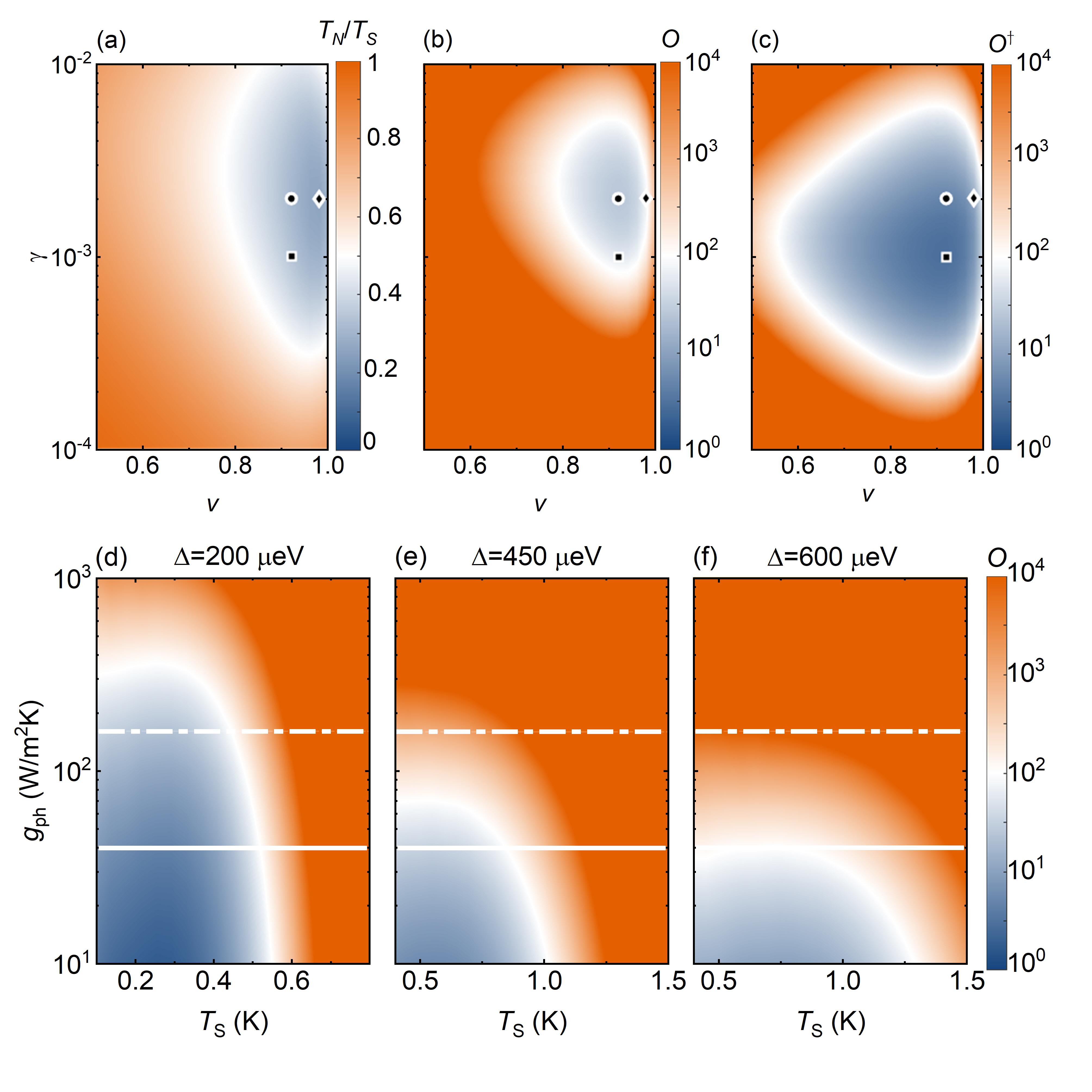}
\caption{\label{fig:o} (a--c) Refrigerator stage with $\Delta=200$~\murm{}eV at $T_S=0.3$~K with $g_\mathrm{ph}=160$~W$/$m$^2$K$^4$ and $A_\gamma=15$~nm$^2$.
Panels (a--c) show $T_N/T_S$, $O$, and $O^\dagger$, respectively, as a function of $\gamma$ and $v$. The minima of $T_N/T_S$, $O$, and $O^\dagger$ are indicated with diamond, circle, and square, respectively.
(d--f) The optimal $O$ with respect to $v$ and $\gamma$ as a function of $g_\mathrm{ph}$ and $T_S$ for refrigeration stages with $\Delta=200$, 450, and 600~\murm{}eV, respectively.
Dash-dotted and solid lines indicate $g_\mathrm{ph}=160$~W$/$m$^2$K$^4$ and 40~W$/$m$^2$K$^4$, respectively.}
\end{figure}

Figures~\ref{fig:o}(a--c) show $T_N/T_S$, $O$, and $O^\dagger$, respectively, as a function of $\gamma$ and $v$ for Si--Al stage at 0.3~K. The minimum of
$O$ and $O^\dagger$ are obtained at smaller $v$ than the minimum of $T_N/T_S$, which shows that it is beneficial to sacrifice some of the cooling power to improve the cooling efficiency. Furthermore, $O^\dagger$ yields a smaller optimum $\gamma$, which is due to the fact that it emphasizes the internal efficiency of the stage, i.e., $P_H$ required to yield a low $T_N$ whereas $O$ optimizes the performance against $P_\mathrm{in}$.

Figures~\ref{fig:o}(d--f) show $O$ for Al-based ($\Delta=200$~\murm{}eV) and V-based ($\Delta=450,600$~\murm{}eV) stages as a function $T_S$ and $g_\mathrm{ph}$. Here $\gamma$ and $v$ have been optimized for each combination of $T_S$ and $g_\mathrm{ph}$ as in Fig.~\ref{fig:o}(b). Regimes with $O \lesssim 10^2$ indicate roughly where Al- and V-based stages yield reasonable performance for practical applications. These results demonstrate how an increased $\Delta$ allows cooling at higher temperatures, but also how phonon heat conductance weakens the performance at higher temperatures. Aluminum can be efficient for $T_S\lesssim 500$~mK, which indicates that vanadium with 3 times higher $\Delta$ is applicable up to $T_S\lesssim 1.5$~K, but only if $g_\mathrm{ph}$ can be made sufficiently low.

\begin{figure}
\includegraphics[width=0.5\textwidth]{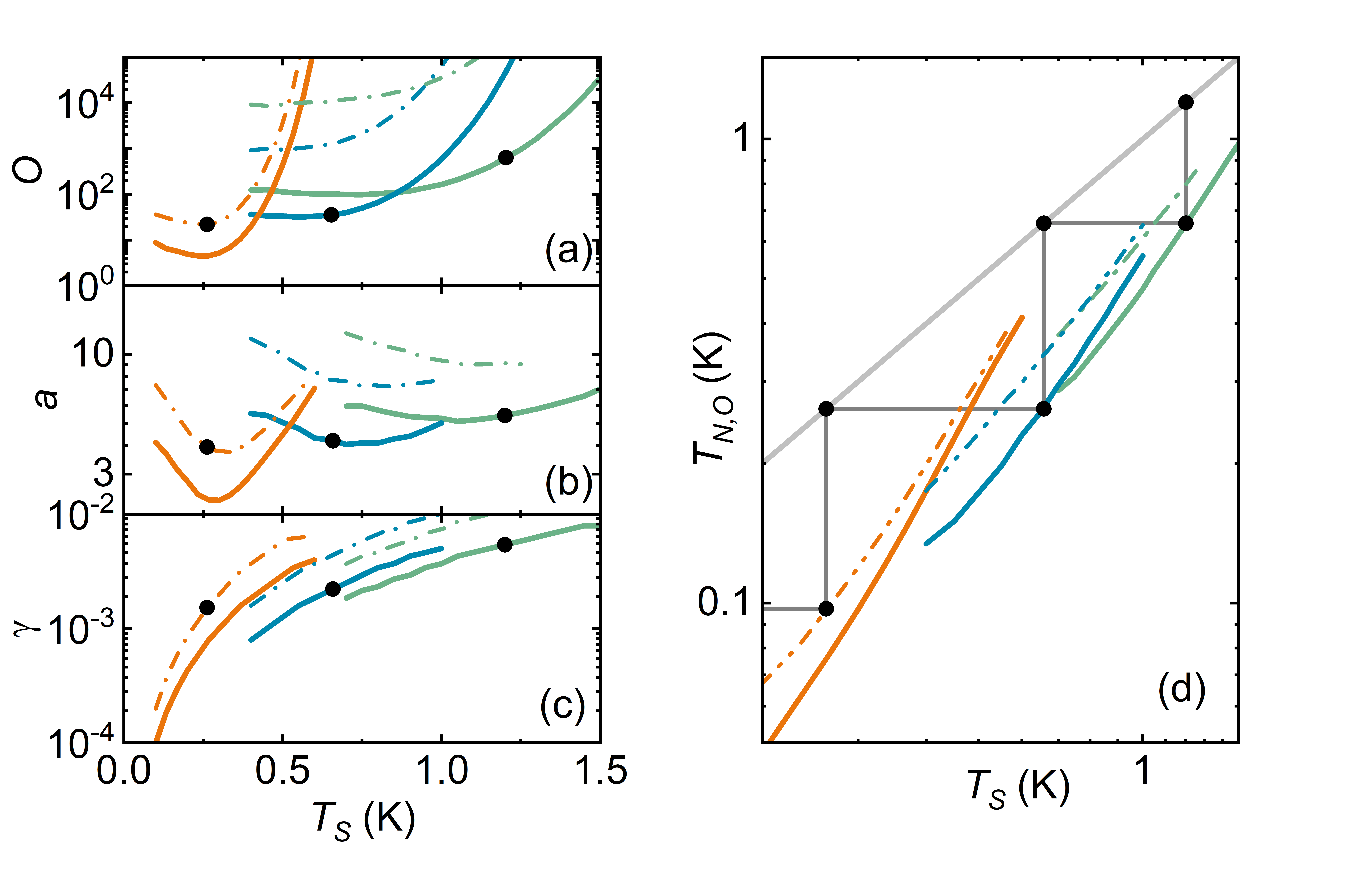}
\caption{\label{fig:results} Results from minimizing $O$ with respect to $v$ and $\gamma$ for refrigeration stages with $\Delta=200$, 450, and 600~\murm{}eV (orange, blue, and green, respectively). Data corresponding to cross-sections indicated with lines in Figs.~\ref{fig:o}(d--f) with $g_\mathrm{ph}=160$~W$/$m$^2$K$^4$ and 40~W$/$m$^2$K$^4$ are shown with dash-dotted and solid lines, respectively.
(a--c) Minimal $O$, and resulting optimal $a$ and $\gamma$, respectively as a function of $T_S$. 
(d) $T_{N,O}$ as a function of $T_S$. The gray line shows the threshold for cooling, i.e., $T_{N,O}=T_S$. The dark gray lines and black circles illustrate expected refrigeration of a cascade with 3 stages starting from $T_0=1.2$~K, where the first, second and third stage have $\Delta=600$~\murm{}eV, $\Delta=450$~\murm{}eV, and $\Delta=200$~\murm{}eV, respectively.
Here, stages 1 and 2 have $g_\mathrm{ph}=40$~W$/$m$^2$K$^4$ and stage 3 has $g_\mathrm{ph}=160$~W$/$m$^2$K$^4$. The parameters of
the 3-stage refrigerator corresponding to the black circles in (d) are shown with similar symbols also in (a--c).}
\end{figure}

Figure~\ref{fig:results} shows outcomes from the optimization along the cross-sections indicated with lines in Figs.~\ref{fig:o}(d--f). Figure~\ref{fig:results}(a) shows that factor 4 improvement of $g_\mathrm{ph}$ can suppress $O$ by factor 100 for a V-based stage, but the difference is smaller for an Al-based stage. This means that the high $P_L$ of high temperature can make refrigeration extremely inefficient. Figure~\ref{fig:results}(b) shows that $a$ is less sensitive than $O$. Inefficient cooling thus yields a small temperature change per stage and increases $F$ rather than $a$. Figure~\ref{fig:results}(c) shows that high $T_S$ and  $g_\mathrm{ph}$ yield high optimal $\gamma$, which is in agreement with Fig.~\ref{fig:gammara} and expected since low $R_A$ maximizes $P_C$. Figure~\ref{fig:results}(d) indicates that cooling from $T_0=1.2$~K down to $T_3=100$~mK, i.e., from $T_0$ of a $^4$He refrigerator is a realistic target for a 3-stage refrigerator if $g_\mathrm{ph}$ can be suppressed below that of Kapitza resistance.

Finally, we evaluate the 3-stage refrigerator by simulating its complete thermal balance that depends on 15 fixed parameters. Table~\ref{tab:balances} shows a comparison of refrigerators A and B, based on minimizing $O$ and $T_N/T_S$ respectively. The parameters of refrigerator B were obtained similarly as explained above, but using $T_N/T_S$ is the optimization criterion instead of $O$. For both A and B, the parameters were obtained iteratively: We first optimized stages 2 and 3 for starting temperatures $T_1$ and $T_2$ obtained as in Fig.~\ref{fig:results}(d). Then we obtained $T_1\ldots T_3$ from the thermal balance and used them to re-optimize stages 2 and 3. Such iteration is not necessary for refrigerator A, since the $O$-based estimate of Fig.~\ref{fig:results}(d) predicts temperatures quite accurately. We used Eq.~(\ref{eq:areas}) to determine $a_n$ also for refrigerator B since non-optimal scaling can yield dramatically worse performance. Note that when the inefficiency of refrigerator B is mitigated with larger $A_n$, it yields almost the same $T_3$ as refrigerator A. In these simulations, we set a reasonable heat load $P_{\mathrm{in},3}/A_3=1$~\murm{}W/mm$^2$, which results in the output power 
$P_{H,1}$ of 50~mW and 240~mW per mm$^2$ of $A_3$ for refrigerators A and B, respectively. This indicates that our method can suppress the energy consumption of the 3-stage refrigerator by about 80\% compared to conventional optimization. For a small $P_{\mathrm{in},3}$, even lower $P_{H,1}$ can be obtained if the last stage is optimized using $O^\dagger$.

\begin{table}
\caption{\label{tab:balances} Parameters and performance of example refrigerators A and B starting from $T_0=1.2$~K with $\gamma$ and $v$ based on minimizing $O$ and $T_N/T_S$, respectively, for each stage separately.}
\begin{ruledtabular}
\begin{tabular}{c  c  a  b  a  b  a  b  a  b}

$n$ & $g_\mathrm{ph}$   & \multicolumn{2}{c}{$A_n/A_3$} & \multicolumn{2}{c}{$\gamma\; (10^{-3})$}    & \multicolumn{2}{c}{$v$}  & \multicolumn{2}{c}{$T$ (mK)}  \\
      & ($\mathrm{W/m^2K^4}$)     &          A  & B           & A & B   &    A & B     & A & B  \\
\hline
1            & 40                    & 32 & 90          & 5 & 5          &  0.88 & 0.94  & 598 & 571   \\
2            & 40                    &  8 & 17           & 2 & 1.6       &  0.94 & 0.98  & 196 & 161  \\
3           & 160                   &    1 & 1          &  0.8 & 0.6   &  0.96 & 0.98  & 89 & 91  \\
\end{tabular}
\end{ruledtabular}
\end{table}


To conclude, we have developed an optimization method for cascaded electronic refrigerators based on NIS or Sm--S tunnel junctions. This method helps to design energy-efficient refrigeration stages that mitigate heating generated in the device. Our results indicate that a practical electrical refrigerator that can cool, e.g., quantum devices from above 1~K to 100~mK is a realistic target, but requires phonon engineering and high-quality tunnel junctions. For the latter goal and for enabling cascaded junction refrigerators we have demonstrated
highly transparent Si–Al and Si–V tunnel junctions with relatively low $\gamma$.

\begin{acknowledgments}
This research was funded by the European Union’s Horizon 2020 research and innovation programme under grant agreements No 766853 EFINED (http://www.efined-h2020.eu/ ), Academy of Finland through projects ETHEC, UQS, and QuMOS (Nos 322580, 310909 and 288907)  and Centre of Excellence program Nos 336817 and 312294.
\end{acknowledgments}

\section*{DATA AVAILABILITY}
Data available on reasonable requests from the authors.

\nocite{*}

\providecommand{\noopsort}[1]{}\providecommand{\singleletter}[1]{#1}%

\end{document}